\documentclass[aps,superscriptaddress,showpacs,amsmath,amssymb,eqsecnum]{revtex4}
\usepackage{color}
\usepackage{graphicx}
\usepackage{bm}
\usepackage{slashed}
\usepackage{ulem}
\usepackage{amsfonts}
\usepackage{graphicx,color}
\usepackage[colorlinks,linkcolor=blue,citecolor=green]{hyperref}
\usepackage{wrapfig}

\makeatletter
\@addtoreset{figure}{section}
\makeatother

\begin{document}

\title{a q-EW-TOPSIS model of grey correlation for supply capacity evaluation}

\author{Jia-Ming Liao}

\author{Yu-Jie Huang}

\author{Ke-Ming Shen}
\email[E-mail: ]{shen$\_$keming.ecut$@$hotmail.com}
\affiliation{School of Science, East China University of Technology,  Nanchang,  China}

\date{\today}

\begin{abstract}

The paper describes a new supply capacity evaluation model based on the non-extensive statistical entropy.  
The traditional EW-TOPSIS model is selected as baseline and the GRA method is used to modify it.  
The correction results in the non-extensive parameter $q$ which leads to the so-called $q$-EW-TOPSIS model.  
This new model has advantages over the traditional EW-TOPSIS model, including the ability to accurately evaluate indicator weights with smaller sample sizes and weaker rules, and a more stable and closer-to-complete structure due to the use of entropy evaluation and mutual restriction between indicators.  
This study provides a more reliable and universal modified EW model. 
It is proved to be a more compatible model with systems and own greater credibility.
\end{abstract}


\maketitle

\section{Introduction}\label{sec:sec1}

Being an advanced way of organization and management technology, modern logistics has attracted more and more attentions of both enterprises and governments.
Nowadays, the methods of logistics capacity evaluation are well studied in various areas \cite{ref1, ref2, ref3, ref4}.
It is also becoming a vital issue to quantitatively analyze the supply capacity, comprehensively examine the advantages and disadvantages from different levels, and comprehensively weigh the advantages and disadvantages, in order to provide a scientific basis for the decision-making of the whole system.

The conventional evaluation method is to evaluate and make decisions through AHP (Analytic Hierarchy Process) \cite{AHP, AHP1}. 
This method conducts a quantitative analysis of qualitative problems and makes the various factors in complex problems orderly by dividing them into interrelated levels. 
The importance of elements in one or two levels is then quantitatively described. 
Finally, the weight factor, which reflects the relevant importance order of elements in each level, is calculated mathematically.
Relatively simple, flexible, and practical, AHP has been widely used as soon as proposed.
However, under the indicators of similar importance, this subjective analysis method cannot solve the defects of large accidental errors existing in different multiple scoring.
In addition, it is prone to errors when there are many evaluation factors and a large scale, and the evaluation results are not accurate and reliable enough.
There are also many classical objective evaluation models, such as the combined evaluation method \cite{cem} and data envelopment method \cite{data}, both of which do not meet the needs of researchers because of their inaccuracy. 
It is often necessary to use a variety of classical algorithms together, and then select the optimal model from them.
In addition, there are a lot of strong adaptive algorithms that can avoid the subjectivity, such as the comprehensive evaluation method of BP (Back-Propagation) neural network \cite{BP,BP1,BP2}.
It carries on the contribution analysis of variables to eliminate insignificant factors, and gives out the current best weight factors by considering the space changed over time.
While BP needs a lot of training samples and long training time as well. 
Moreover, the complexity of this algorithm is quite high.
On the other hand, for most traditional algorithms, being a class of static models, it is also difficult to maintain the quality of the model's accuracy as data evolves.
Furthermore, when the data size is not sufficient and the impact of the loss of information during the acquisition channel cannot be ignored, there is no corresponding method to compensate for this loss of information, which mostly leads to large changes in the evaluated weights with data increasing.

TOPSIS (Technique for Order Preference by Similarity to Ideal Solution) \cite{TOPSIS, TOPSIS0} is an effective evaluation method based on multi-objective decision-making. 
It normalizes the original data, calculates the weight of limited evaluation objects by weighting, and evaluates the quality of the evaluation objects.
EW-TOPSIS method (Entropy weight TOPSIS method) \cite{EWM}, as a modified TOPSIS method with weights determined by information entropy, ensures the accuracy of evaluation.
However, considering the uncertainty of information with data changing, the complexity of the model is moderate. 
It somehow makes the determined weights change greatly, which leads to the fact that stable weights should be reached around the EW-TOPSIS results rather than themselves. 
Thus the data need to be perturbed correspondingly many times and this model is also to be synthesized many times to find a metastable point, which is then cumbersome. 

In order to further optimize the TOPSIS model, we hereby introduce the non-extensive statistical theory.
It starts by generalizing the standard Maxwell-Boltzmann statistics and covers lots of entropic forms such as the ones proposed by Renyi \cite{Renyi}, Tsallis \cite{TA}, Abe \cite{Abe}, Kaniadakis \cite{Kaniadakis} among others.
In this work, the Tsallis entropy is applied which obtains a non-extensive parameter, $q$.
When $q \to 1$ it recovers the classical theory.
Based on the non-extensive Tsallis entropy weight, a $q$-EW-TOPSIS model is then applied into the investigations on the supply capacity evaluation.

\section{theoretical framework}\label{sec:sec2}

\begin{figure}[htb]
\centering
\includegraphics[width =0.7\linewidth]{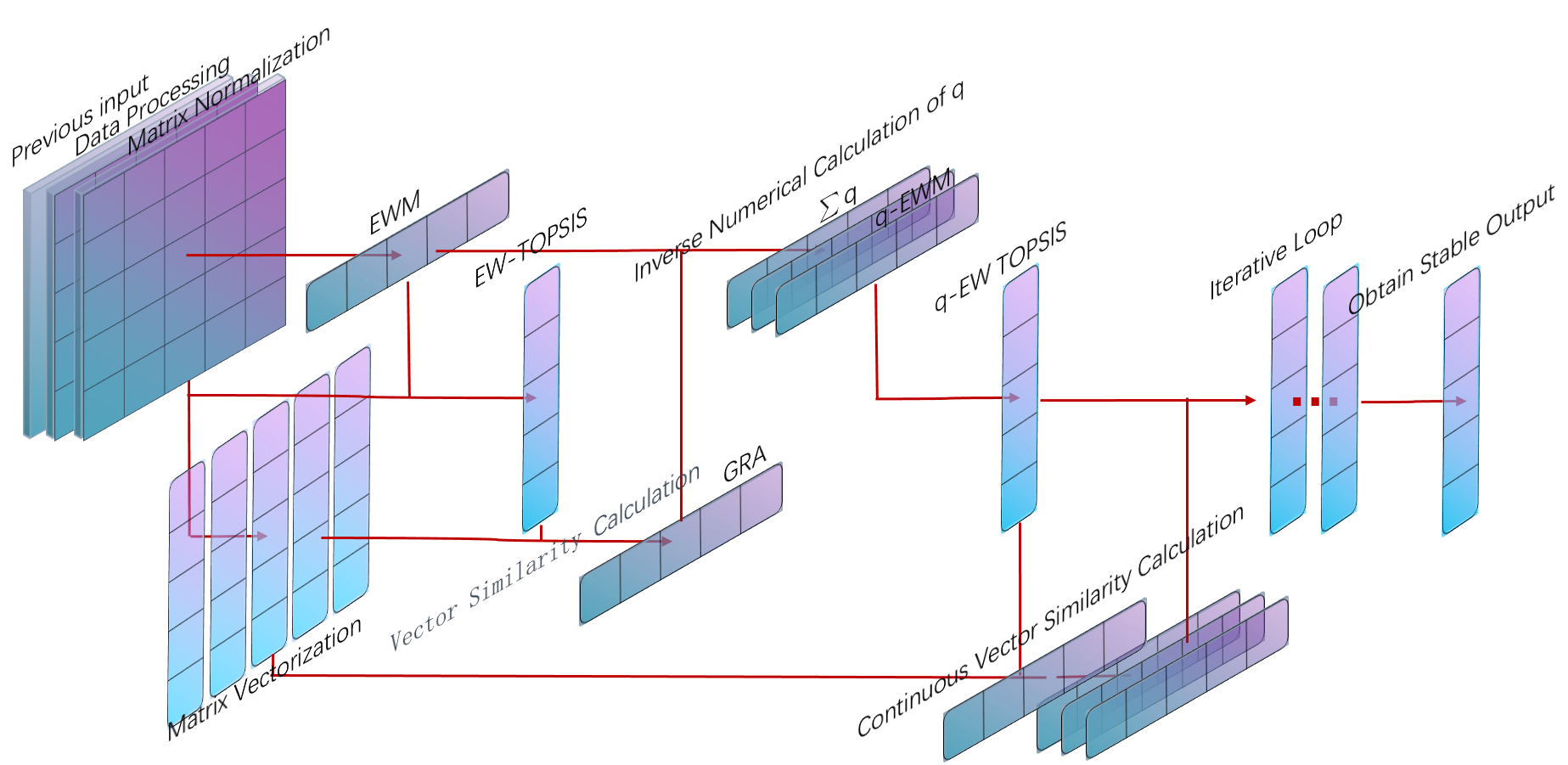}
\caption{The algorithm flow chart of the q-EW TOPSIS model.}
\label{chart}
\end{figure}

We have created an algorithmic flowchart to facilitate reader comprehension, as illustrated in Figure \ref{chart}.

Subsequently, I will delve into the specific details of the algorithm.

In this paper, we utilize partially available data provided by the Mathematical Modeling Suppliers and Customers of the 
Chinese Society of Industrial and Applied Mathematics to verify the advantages of the model and algorithm, as the model 
and algorithm do not strictly require specific data size and form \cite{dataset}.

\subsection{EW-TOPSIS model}

TOPSIS method was firstly proposed by C. L. Hwang and K. Yoon in 1981 \cite{ref6}. 
It ranks a limited number of evaluation objects according to their proximity to the idealized target, and evaluates the relative merits of the existed objects.
It is a multi-index comprehensive evaluation method, which can synthesize multiple indicators into a single matrix for investigation.
In our model, the EW-TOPSIS model is applied to calculate out a reference vector, which serves as a set of evaluation sequences to be evaluated; the scores are calculated from original weights under ideal conditions, namely, the scores of different suppliers in this work.

First of all, all the relevant parameters are specifically classified as follows.
The supply stability is defined as a minimum indicator which is also called the cost-type indicator, namely the smaller value of it leads to the better result.
On the contrary, there are three maximum indicators, the supply quantity (i.e., efficiency-type indicator), the supply continuity and the ambiguous supply capacity.
All of them should be forward normalized in  matrixes by using the range method next.

For the minimum indicators, 
\begin{equation}
x'_{i j}=\frac{x_{\max}-x_{j }}{x_{\max }-x_{\min }},
\end{equation}
where $x'_{ij}$ stands for the positively normalized indicators, $x_{min(max)}$ for its minimum(maximum) value of $j$-items.
Similarly, for maximum indicators,
\begin{equation}
x'_{i j}=\frac{x_{j}-x_{\min }}{x_{\max }-x_{\min }}.
\end{equation}
Thus can we obtain the probability matrix as,
\begin{equation}
p_{i j}=\frac{x'_{i j}}{\sum_{i}^{m} x'_{i j}} \quad, \quad 0 \leq p_{i j} \leq 1,
\label{equa-pij}
\end{equation}
with $m$ denoting the total number of microstates of system.

To better classify the EW-TOPSIS model, we firstly consider the classical statistics.
It is well known that Shannon entropy goes like,
\begin{equation}
S_{Sh}[p_i]=E\left[\ln\frac{1}{p_i}\right]=-\sum_{i=1}^{m} p_{i} \ln p_{i},
\label{entro-Sh}
\end{equation}
where $E[x]$ represents the expected value of a variable $x$.
Worthy to note that hereby the Boltzmann constant $k_B=1$ is applied without losing generality.
The information entropy $e_j$ of the $j$th-parameter is then defined as
\begin{equation}
e_j=S_{j}[p_{ij}]=-\sum_{i=1}^{m} p_{i j} \ln \left(p_{i j}\right).
\label{equa-ej}
\end{equation}
The information utility value $d_j$ is calculated as follows:
\begin{equation}
d_j=1-e_{j},
\label{equa-dj}
\end{equation}
together with the weight factor for each evaluation indicator:
\begin{equation}
W_S(j)=\frac{d_j}{\sum_{j=1}^n d_j},
\label{equa-Wsj}
\end{equation}
where $n$ denotes the number of $j$th-evaluation indicator.

In order to eliminate the influence of dimensions of different indexes, the matrix needs to be further standardized.
Next, we normalize the forward matrix X, with its element $z_{ij}$
\begin{equation}
z_{i j} =\frac{x_{i j}}{\sqrt{\sum_{i=1}^{m} x_{i j}^{2}}}
\label{equa-zij}
\end{equation}
and $z^*_{ij}$
\begin{equation}
z_{i j}^{*} =z_{i j} \cdot W_S(j).
\label{equa-zsij}
\end{equation}
The maximum or optimized scheme is then defined as:
\begin{flalign}
Z^{*+}&=(Z_1^{*+},~Z_2^{*+}, ~\cdots, ~Z_m^{*+}) \nonumber \\
&=\big( max\{ z_{11}, z_{21}, \cdots, z_{n1} \},~max\{ z_{12}, z_{22}, \cdots, z_{n2} \},~\cdots, ~max\{ z_{1m}, z_{2m}, \cdots, z_{nm} \} \big).
\label{equa-ZS+}
\end{flalign}
Similarly we have the minimum or deteriorated scheme,
\begin{flalign}
Z^{*-}&=(Z_1^{*-},~Z_2^{*-}, ~\cdots, ~Z_m^{*-}) \nonumber \\
&=\big( min\{ z_{11}, z_{21}, \cdots, z_{n1} \},~min\{ z_{12}, z_{22}, \cdots, z_{n2} \},~\cdots, ~min\{ z_{1m}, z_{2m}, \cdots, z_{nm} \}  \big).
\label{equa-ZS-}
\end{flalign}
Using them the distance between the $i$th-evaluation subject ($i=1, 2, \cdots, n$) and the maximum (minimum) value is well clarified,
\begin{equation}
D_{i}^{\pm}=\sqrt{\sum_{j=1}^{m}\left(Z_{j}^{*\pm}-z_{i j}^{*}\right)^{2}}.
\label{equa-Di}
\end{equation}
The normalized score $S_{i}$ for the $i$th subject is finally obtained as:
\begin{equation}
S_{i}=\frac{D_{i}^{-}}{D_{i}^{+}+D_{i}^{-}}.
\label{equa-Si}
\end{equation}
It is obvious that $S_i$ belongs to [0,1], and larger values of $S_i$ lead to the larger $D_i^+$, which gets closer to the maximum scheme, otherwise the opposite.
Thus can we obtain a set of reference vectors calculated using the original EW-TOPSIS model as the scores.

\subsection{Grey Relational Analysis}

Before utilizing the non-extensive theory to generalize the classical EW-TOPSIS model, it is crucial to thoroughly investigate the correlation degree among the evaluation sequences and the reference scores, in order to iteratively refine the data.
The concept of grey systems was firstly proposed by Professor Deng Jilong \cite{GRA}, a control scientist and engineer, in comparison to white systems and black systems. 
According to control theory, the color usually represents the amount of information known about a system.
White systems represent systems with sufficient information, such as a mechanical system, where the relationships between elements are determinable. 
On the other hand, black systems represent systems where the structure is not clear, often referred to as black boxes. 
Grey systems fall in between, representing systems where only partial understanding is available.
In our model, due to the lack of information and the impact of the loss of information, the correlation degree between the evaluation sequences and the reference scores can vary, and thus the use of grey systems theory to correct the deviation is appropriate.

The Grey Relational Analysis (GRA) is a very active branch of the grey systems theory. 
Its basic principle is to determine the proximity of the relationship between different sequences based on the geometric shape of the corresponding sequence curves. 
Many researchers have already combined this method with other algorithms, validating its effectiveness in supplier selection and evaluation \cite{GRA1,GRA2}. 
Therefore, GRA will be applied in this study to calculate the correlation degree among the evaluation sequences and the reference scores.

GRA is to calculate the  degree in this work since in the grey system grading curves of each parameter could be quantitatively analyzed in it.
Nevertheless, it is only adopted to obtain the correction amplitude to modify the idealized EW-TOPSIS model because of the inevitable errors existed in the evaluation of weight factors.

Firstly consider the characteristic sequence of system:
\begin{equation}
R=\left[\begin{array}{ccc}
x_{1}(1), x_{1}(2), \cdots, x_{1}(n), S_{1} \\
x_{2}(1), x_{2}(2), \cdots, x_{2}(n), S_{2} \\
\vdots  \qquad \vdots \qquad \ddots\qquad \vdots\\
x_{m}(1), x_{m}(2), \cdots, x_{m}(n), S_{m}
\end{array}\right],
\label{equa-css}
\end{equation}
where the subsequences $x_i(j),~(i=1,2,\cdots, m; j=1,2,\cdots, n)$ correspond to the values of $j$th-indicators with $i$th-ID.
The score $S_i$ is following the definition as in Eq.(\ref{equa-Si}).

The correlation coefficients of the subsequence are then calculated for each sequence.
The correlation coefficient on $S_i$ with $x_i$ at $j$ point is defined as:
\begin{equation}
\gamma\left(S_{i}, x_{i}(j)\right)=\frac{\min _{i} \min _{j}\left|S_{i}-x_{i}(j)\right|+\xi \max _{i} \max _{j}\left|S_{i}-x_{i}(j)\right|}{\left|S_{i}-x_{i}(j)\right|+\xi \max _{i} \max _{j}\left|S_{i}-x_{i}(j)\right|}.
\label{equa-gamma1}
\end{equation}
Here $\min _{i} \min _{j}\left|S_i-x_{i}(j)\right|$ denotes the absolute value of the  secondary minimum difference and $\max _{i} \max _{j}\left|S_i-x_{i}(j)\right|$ of the maximum one, respectively.
On the other hand, $\xi$ is introduced as the grey distinguish coefficient with its range of $[0,1]$; it is shown better when $\xi<0.5468$.

Using Eq.(\ref{equa-gamma1}) another coefficient between $X_i$ and $X$ could be written as:
\begin{equation}
\gamma\left(S(j), X_{i}\right)=\frac{1}{m} \sum_{i=1}^{m} \gamma\left(S_{i}, x_{i}(j)\right).
\label{equa-gamma2}
\end{equation}
Since the grey correlation coefficients are dimensionless and the contribution of a group of vectors to the grey system could be quantitatively analyzed under the same grey resolution coefficient, they are normalized to obtain the contribution weight of each index to the system\cite{ref8},
\begin{equation}
W_{r}(j)=\frac{\gamma_{j}}{\sum_{j=1}^{n} \gamma_{j}}.
\label{equa-Wrj}
\end{equation}

In the analysis of the supply ability score in this work, $m=402$ indicates the number of suppliers and $n=4$ is for four indicators after the forward transformation.
The scoring matrix $S$ obtained by the EW-TOPSIS model is used as a reference sequence of $S_i$ in this work and other indicators as comparison sequences of $x_{i}(j)$. 
Furthermore, the correlation degree is calculated through four indicators, i.e., the supply stability, quantity, continuity and the ambiguous supply capacity.
The weight curve then gives a guide to how much the model should change (up or down).

\subsection{$q$-EW-TOPSIS model}

It is inevitable that there is a significant amount of noise in the process of information acquisition, which deviates significantly from the ideal case. 
When the data volume is not large enough to ignore it, all simulations are no longer exactly done by classical statistical models. 
In this subsection, a non-extensive statistical theory is nicely applied as a useful tool for further simulating the entropy weight method above.

The Tsallis non-extensive statistics, as a widely-used generalization of classical Boltzmann-Gibbs statistics, introduces a non-extensive quantity $q$ to modify it with a noise loss. 
When $q\to 1$ it returns to the classical situation.

Specifically, in 1988 C. Tsallis firstly proposed the non-extensive statistics \cite{TA}.
Although it is one of generalizations of the classical entropy, c.f. Eq.(\ref{entro-Sh}), the Tsallis entropy
\begin{equation}
S_{q}=\ln _{q} W \equiv \frac{W^{1-q}-1}{1-q}
\label{entro-Sq}
\end{equation}
has been nicely applied into various kinds of fields.
It is also often seen as the form of probabilities,
\begin{eqnarray}
S_{Ts}=E[\ln_q(\frac{1}{p_i})]=\frac{\sum_i^m p_i^q-1}{1-q}.
\label{entro-STs}
\end{eqnarray}
Note that here the $q$-logarithm is introduced, 
\begin{eqnarray}
\ln_q(x):=\frac{x^{1-q}-1}{1-q} \quad (x>0, q\neq 1),
\label{equa-qln}
\end{eqnarray}
with its inverse function, the $q$-exponential function,
\begin{eqnarray}
\exp_q(x):=[1+(1-q)x]^{\frac{1}{1-q}} \quad (q\neq 1).
\label{equa-qex}
\end{eqnarray}
Both of them will go back to the normal case when the limit of $q\to 1$ is taken.
Moreover, the most obvious feature of non-extensive entropy is that the entropy of a system consisting of two subsystems is not a direct addition of the entropic functions of the two subsystems, namely the extensive property of the entropy function is no longer preserved.
With this non-extensive parameter, this theory could be more widely and successfully used in solving problems where the traditional statistical models could not be applied.
For example, it has been proved to be successful in the fields of self-gravitation systems, neural networks, turbulence, etc..

In this paper, we introduce the non-extensive theory in order to modify the values of entropy weight factors by re-fitting the entropy weight curves within kinds of similar algorithms.
A non-idealized model will be then constructed, which eventually covers the complex connections in systems.
This will also greatly improve the model's noise immunity, increase the model's accuracy and reliability during data iteration, and make the model more generalizable.

Similar to the classical case in Eq.(\ref{equa-ej}), the generalized formula of information entropy value can be deduced as
\begin{equation}
\tilde{e}_j=S_{T s}\left(p_{ij}\right)=\frac{\sum_{i}^{m} p_{ij}^{q}-1}{1-q}.
\label{equa-eqj}
\end{equation}
By solving the constraint equation
\begin{equation}
\tilde{e}_j = W_{r}(j)
\label{equa-qeW}
\end{equation}
we could give out the selected $q$ values of different evaluation sequence in the $q$-entropy weight method.
Note that $W_r(j)$, shown in Eq.(\ref{equa-W}), represents the grey correlation correction weight.
At the same time, for consistency it is considered that $q$ will be unity only when the grey correlation sequence cannot give out an exact $q$.
Finally, the expected $q$-value is re-calculated as:
\begin{equation}
q=\frac{\sum_{j=1}^{n} q_{j}}{n},
\label{equa-qall}
\end{equation}
and the normalized weights with the $q$ above are given below:
\begin{equation}
W_{T}(j)=\frac{\tilde{e}_j}{\sum_{j=1}^{n} \tilde{e}_j}
\label{equa-Wtj}
\end{equation}

\section{Results and Discussions}\label{sec:sec3}

\subsection{Data Processing}

Data were taken within five years when the number of suppliers is the most stable for processing. 
And the four evaluation indicators of supply capacity are listed hereafter.
\begin{enumerate}
\item Supply stability

Supply stability is the variance value $D(X)$ of the total raw materials provided by each supplier within five years, 
\begin{equation}
D\left(X_{i}\right)=E\left[ X_{i}^{2}\right]-\mu_{i}^{2},
\label{equa-DX}
\end{equation}
where $E[X_i^2 ]$ is the mean square value of raw materials provided by a single supplier in the past five years, and $\mu_i^2$ is the square of the mean value of raw materials provided by a single supplier in the past five years.

Data matrix {\bf X} is obtained by the imported data sets and then transposed.
$D(X_i)$ of each column of this generating matrix is calculated to get the single-row matrix {\bf EV}.
It concludes nearly all the information on the supply stability from every supplier, so that it will be defined as one indicator of the supply stability correspondingly. 
Last not least, an empty matrix {\bf W} is also constructed and denoted as the original evaluation matrix with loading the matrix {\bf EV}.

\item Supply quantity

Summing the matrix {\bf X} row by row, we can obtain a single-row vector {\bf S} about the supply quantity of suppliers,
\begin{equation}
S=\sum_{j=1}^{n} X_{i j}.
\label{equa-Supplyquantity}
\end{equation}
As another evaluation indicator, {\bf S} will be also loaded into the matrix {\bf W}.

\item Supply continuity

For indicating the supply continuity, we re-build one empty matrix comparing with the matrix {\bf X}.
The number of 0 is for no supply of the current month while 1 is the opposite.
Then do the re-summation column by column and obtain the single-row matrix {\bf L}.
Being the indicator of evaluating the supply continuity, it is then loaded into {\bf W}.

\item Ambiguous supply capacity

Furthermore, two different matrixes will be generated by MATLAB by subtracting the data sets of both the enterprise's order quantity and  supplier's supply quantity.
For the indicator matrix of ambiguous supply capacity, the difference of these two should be made and re-summed column by column next.
The final single-row matrix {\bf T}, as our last indicator, will be loaded into the matrix {\bf W} as well.

\end{enumerate}

\subsection{Data analysis}
Relevant results are graphically shown in this part.
Firstly the original EW-TOPSIS model was considered and its normalized score table of $S_i$ (seen as in Eq.(\ref{equa-Si})) for the $i$th ID of suppliers was given as shown in Table \ref{table-Si}.
For simplicity only the first 48 of all 402 suppliers are listed, which will be as our reference standards for comparisons between the original and non-extensive EW-TOPSIS models.

\begin{table}[htb]
\caption{Normalized Scores in Eq.(\ref{equa-Si}) of the first 48 of all 402 suppliers:}
\scalebox{1.3}[1.3]{
\begin{tabular}{c|c|c|c|c|c|c|c|c|c|c|c}
\hline
\hline
S/N & ID & Score & S/N & ID & Score &S/N & ID & Score &S/N & ID & Score \\
\hline
1 & 229 & 99.997 & 13 & 330 & 41.883 & 25 & 201 & 22.650 & 37 & 037 & 19.030 \\
2 & 361 & 92.635 & 14 & 356 & 40.331 & 26 & 365 & 21.991 & 38 & 218 & 18.963 \\
3 & 140 & 85.343 & 15 & 268 & 40.204 & 27 & 031 & 21.931 & 39 & 007 & 18.353 \\
4 & 108 & 68.966 & 16 & 306 & 39.311 & 28 & 395 & 21.848 & 40 & 266 & 18.328 \\
5 & 151 & 56.638 & 17 & 194 & 33.519 & 29 & 040 & 20.674 & 41 & 123 & 18.323 \\
6 & 340 & 50.647 & 18 & 352 & 30.792 & 30 & 364 & 20.294 & 42 & 338 & 18.278 \\
7 & 282 & 50.111 & 19 & 348 & 29.833 & 31 & 367 & 20.017 & 43 & 150 & 18.080 \\
8 & 275 & 47.359 & 20 & 143 & 29.464 & 32 & 055 & 19.769 & 44 & 114 & 17.923 \\
9 & 329 & 46.843 & 21 & 307 & 25.659 & 33 & 346 & 19.686 & 45 & 314 & 17.763 \\
10 & 139 & 45.278 & 22 & 247 & 24.418 & 34 & 080 & 19.293 & 46 & 291 & 17.087 \\
11 & 131 & 42.095 & 23 & 374 & 23.162 & 35 & 294 & 19.257 & 47 & 086 & 17.028 \\
12 & 308 & 41.968 & 24 & 284 & 22.745 & 36 & 244 & 19.041 & 48 & 098 & 15.855 \\ 
\hline
\hline
\end{tabular}
}
\label{table-Si}
\end{table}

\begin{figure}[htb]
\centering
\includegraphics[width =0.7\linewidth]{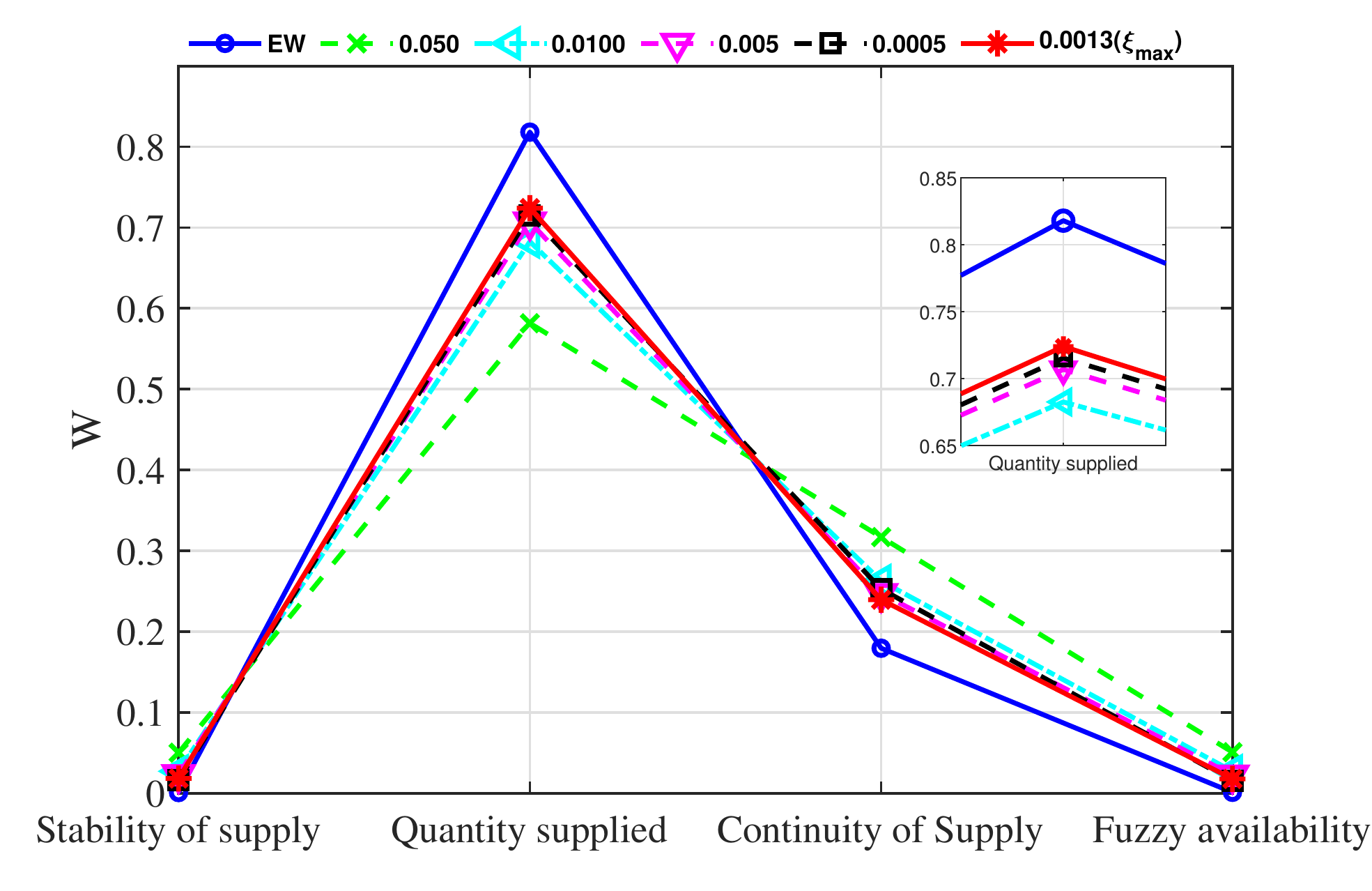}
\caption{The graph of grey correlation coefficient weight and original weight of each index under different gray discrimination coefficient ($\xi$) }
\label{plot-W}
\end{figure}

Next we go on looking into the characteristic sequence of system as well as its contribution weight.
Figure \ref{plot-W} describes the weight distributions with the four indicators above.
Being the baseline, the result of classical EW model, as in Eq.(\ref{equa-Wsj}), is firstly present as the blue dotted line.
All the other weights are derived by Eq.(\ref{equa-Wrj}) with different values of the gray discrimination coefficient $\xi$.

In this iteration, we will start with a gray discrimination coefficient of 0.001, with a step size of 0.0001, up to 0.005. 
We will then select the gray discrimination coefficient that is most similar to the original entropy weight as the optimal coefficient for this iteration and incorporate it into future revisions.

It is obvious to see that the GRA curve with $\xi=0.0013$, namely the red dashed line, best fits the classical EW curve among all.
Corresponding correlation coefficients $\gamma(S_i, x_i(j))$ are then calculated out, based on Eq.(\ref{equa-gamma1}) and Eq.(\ref{equa-gamma2}).
Finally the GRA weight factors with this $\xi$ is given out.

In order to clarify the difference between the classical and the $q$-EW-TOPSIS models, the non-extensive normalized weights (seen in Eq.(\ref{equa-Wtj})) should be studied too.
GRA weight factors above for all four indicators are substituted into the relation of Eq.(\ref{equa-qeW}) to get the non-extensive parameters $q_i$, respectively, which are clearly collected in Table \ref{table-q}.
Note that from now on the average non-extensive parameter $q=\sum_j^n q_j /n$ is introduced without losing of generality. 
In virtue of the validity of weight, the weight factor in the $q$-EW-TOPSIS model is eventually solved, following Eq.(\ref{equa-Wtj}). 

\begin{table}[htb]
\caption{Non-extensive parameters of $q_j$ calculated by Eq.(\ref{equa-qeW}) for each indicator:}
\scalebox{1.3}[1.3]{
\begin{tabular}{c|c|c|c|c|c}
\hline
\hline
Indicators & SS & SQ & SC & ASC & Average value\\
\hline
Parameter $q$ & 0.3216 & 0.1346 & 1.0001 & 0.0993 & 0.3888 \\ 
\hline
\hline
\end{tabular}
}
\label{table-q}
\end{table}

\begin{figure}[h]
\centering
\caption{QEWM-TOPSIS score of each index}
\includegraphics[width =0.7\linewidth]{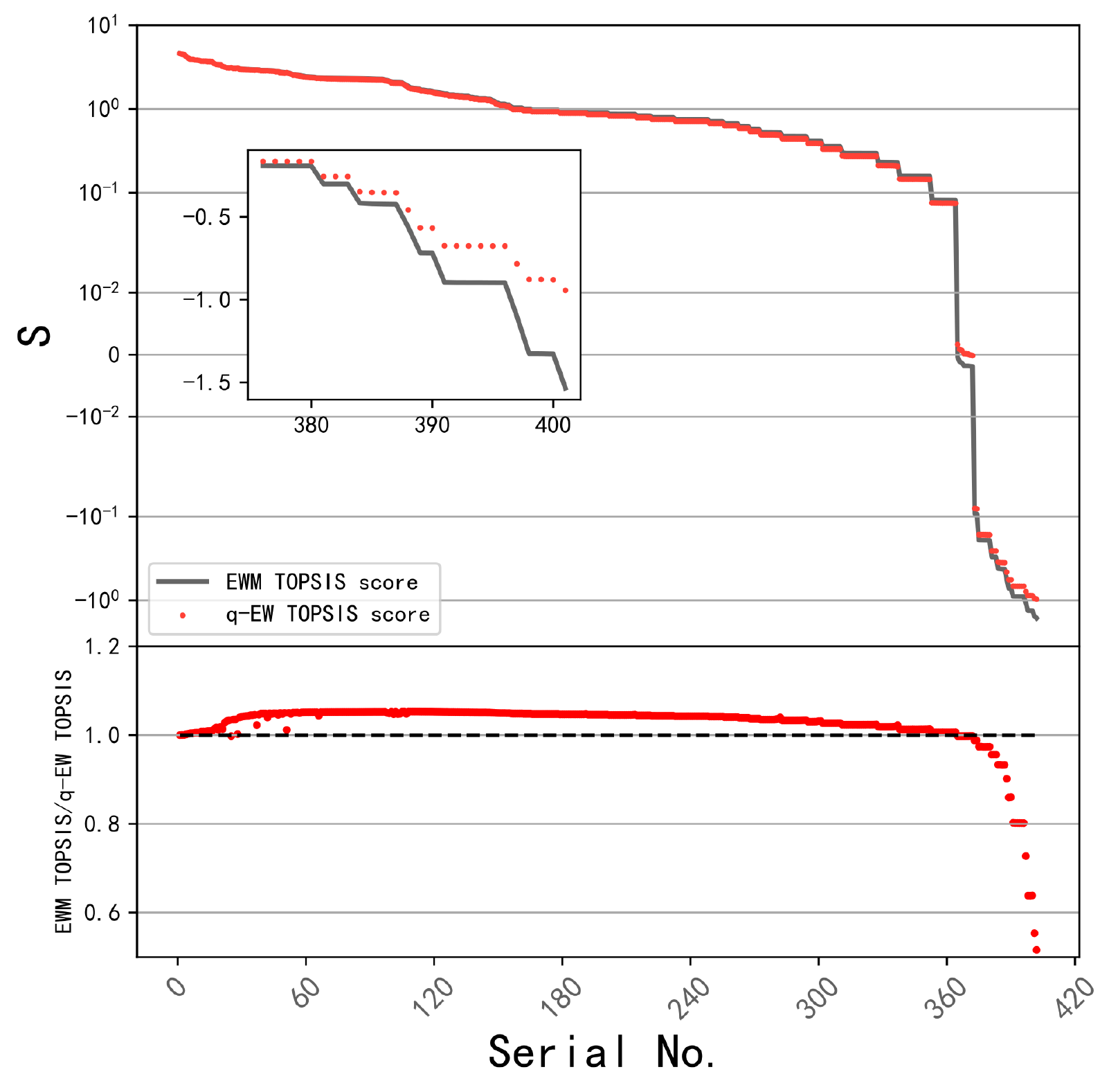}
\label{plot-S}
\end{figure}

\begin{table}[!h]
\caption{Non-Extensive Scores of the first 48 of all 402 suppliers in $q$-EW-TOPSIS model:}
    \scalebox{1.3}[1.3]{
    \begin{tabular}{c|c|c|c|c|c|c|c|c|c|c|c}
    \hline
    \hline
    S/N & ID & Score & S/N & ID & Score &S/N & ID & Score &S/N & ID & Score \\
    \hline
    1 & 229 & 99.994 & 13 & 330 & 41.553 & 25 & 284 & 22.016 & 37 & 244 & 18.204 \\
    2 & 361 & 92.615 & 14 & 356 & 39.980 & 26 & 395 & 21.800 & 38 & 218 & 18.125 \\
    3 & 140 & 85.316 & 15 & 268 & 39.852 & 27 & 365 & 21.240 & 39 & 338 & 17.582 \\
    4 & 108 & 68.855 & 16 & 306 & 38.946 & 28 & 031 & 21.177 & 40 & 007 & 17.509 \\
    5 & 151 & 56.451 & 17 & 194 & 33.059 & 29 & 040 & 19.880 & 41 & 266 & 17.484 \\
    6 & 340 & 50.411 & 18 & 352 & 30.277 & 30 & 364 & 19.489 & 42 & 123 & 17.480 \\
    7 & 282 & 49.870 & 19 & 348 & 29.470 & 31 & 367 & 19.204 & 43 & 150 & 17.239 \\
    8 & 275 & 47.091 & 20 & 143 & 28.919 & 32 & 55 & 18.949 & 44 & 114 & 17.107 \\
    9 & 329 & 46.570 & 21 & 307 & 25.313 & 33 & 346 & 18.863 & 45 & 314 & 16.934 \\
    10 & 139 & 45.028 & 22 & 247 & 23.741 & 34 & 037 & 18.607 & 46 & 291 & 16.294 \\
    11 & 131 & 41.767 & 23 & 201 & 22.697 & 35 & 080 & 18.461 & 47 & 086 & 16.283 \\
    12 & 308 & 41.639 & 24 & 374 & 22.446 & 36 & 294 & 18.424 & 48 & 098 & 15.100 \\
    \hline
    \hline
    \end{tabular}
    }
    \label{table-Siq}
\end{table}

By incorporating the non-extensive normalized weights into Eq.(\ref{equa-eqj}, \ref{equa-Wtj}), the new model's score sequence can be obtained. We will then order 
it and present the top 48 in the table.

It is of great importance to research on the comparisons between these two different models.
As shown in Figure \ref{plot-S}, the scores of all suppliers within these two different models are clearly compared.
There are no big differences among them all, especially the sequence ordered by scores.
For instance, the best 10 suppliers share the same ID numbers within both of models although they own different scores.
This is especially indicated when comparing the data results in Table \ref{table-Si} and Table \ref{table-Siq}.
For deeper clarification, Table \ref{table-Siq} still exhibits the top 48 suppliers with their scores.

In the above figure, the upper part shows the distribution of scores for two evaluation sequences, and the lower figure shows the error analysis of the original score and the $q$-EW TOPSIS score (measured by the ratio method). 
It can be clearly seen that after reordering the supplier evaluation ID according to the original EW-TOPSIS model score, the old model and the new model score curve are compared. 
It can be seen that a certain decrease in the score of the middle part of the sequence is made, and a certain compensation is made for the score of the tail part of the sequence. 
Overall, the score fluctuates but is not inconsistent with the facts, which proves that the $q$-EW-TOPSIS model does not deviate too much from the original model's score distribution and verifies the reliability of the model. 
On the other hand, using the new model for scoring, 99 suppliers' rankings have changed, with an impact rate of 24.63$\%$. 
Therefore, in terms of supplier selection, this method is also different from traditional evaluation methods. 
This ensures the correctness and reliability of the model itself.

In order to better review this model, the difference proportion $\delta$ is established as
\begin{equation}
\delta(j)=\frac{\left|W_S(j)-W_T(j)\right|}{W_T(j)}.
\label{equa-delta}
\end{equation}

In this paper we have the values of it as 0.5352, 0.0096, 0.0606 , and 0.4741, for four evaluation indicators, respectively. 
Normally the smaller value of it leads to the greater influence of non-extensivity on the system.
Its average value is then given out, $\bar{\delta}=0.2396$.

In order to validate the model's robustness and generalization capabilities, we need to artificially manipulate the data to simulate the impact of data changes on model weights under real conditions, in order to evaluate the quality of our model.

This further demonstrates the robustness and generalizability of the model, as it is able to maintain a stable weight distribution even when the amount of data changes. 
Additionally, the small variance in the weight averages suggests that the model is able to provide consistent results under different conditions. 
Overall, this experiment shows that the model is able to perform well in dynamic and unpredictable environments.

\begin{table}[htb]
\caption{The mean and variance of the weight of the original weight w0 and the weight of this model w1 during the data iteration process.}
\scalebox{1.3}[1.3]{
\begin{tabular}{c|c|c|c|c}
\hline
\hline
Indicators & SS & SQ & SC & ASC \\
\hline
$w0_{ave}$ & 0.0057 & 0.7733 & 0.2135 & 0.0075 \\ 
\hline
$w1_{ave}$ & 0.0064 & 0.7724 & 0.2130 & 0.0082 \\ 
\hline
$w0_{var}$ & 0.0000 & 0.0031 & 0.0032 & 0.0075 \\ 
\hline
$w1_{var}$ & 0.0000 & 0.0024 & 0.0025 & 0.0000 \\ 
\hline
\hline
\end{tabular}
}
\label{table-allw}
\end{table}

\begin{figure}[!h]
    \centering
    \includegraphics[width =0.7\linewidth]{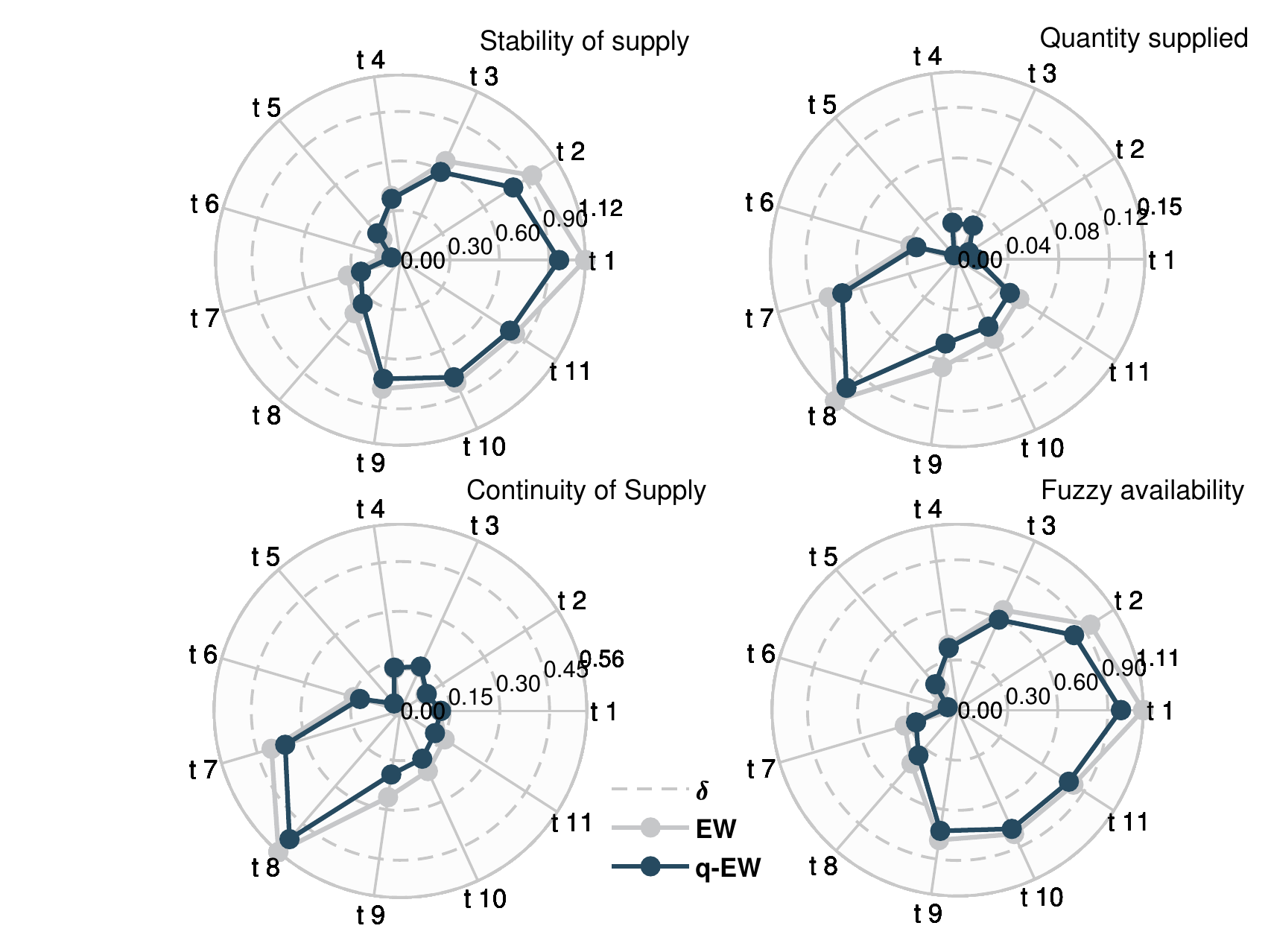}
    \caption{Radar chart of the weights of the corresponding evaluation index by taking the number of different suppliers as the step size.}
    \label{plot-err-radar}
\end{figure}

Table \ref{table-allw} illustrates the average and variance of the original weight $w0$ and the weight of the proposed model $w1$ throughout the entire data iteration process. 
As shown, the mean of the weights for the proposed model is relatively similar to that of the original model, however, the variance of the weights for the proposed model is slightly better than the original model. 
To better visualize the improvements of the proposed model, next the relative error of each evaluation indicator is shown at each data update.

Next we set the average weight of each model after each iteration as the standard weight, and plot the relative error of the weight (current period weight - standard weight / standard weight) as a radar chart, seen in Figure \ref{plot-err-radar}. 
As can be seen from the chart, the weights determined by the new method have errors that are generally smaller than those of the original model under general conditions.

In summary, we have demonstrated the superiority of the proposed non-extensive statistical theory in simulating the entropy weight method. 
By comparing the relative errors of the weights calculated by the original and present models, we have shown that the new-proposed model has better robustness and generalization ability. 
The results of our study have verified the reliability and effectiveness of the proposed model in mitigating the impact of noise and incomplete information in data iteration, making it able to maintain high accuracy and credibility under different data conditions.

\begin{figure}[!h]
\centering
\includegraphics[width =0.7\linewidth]{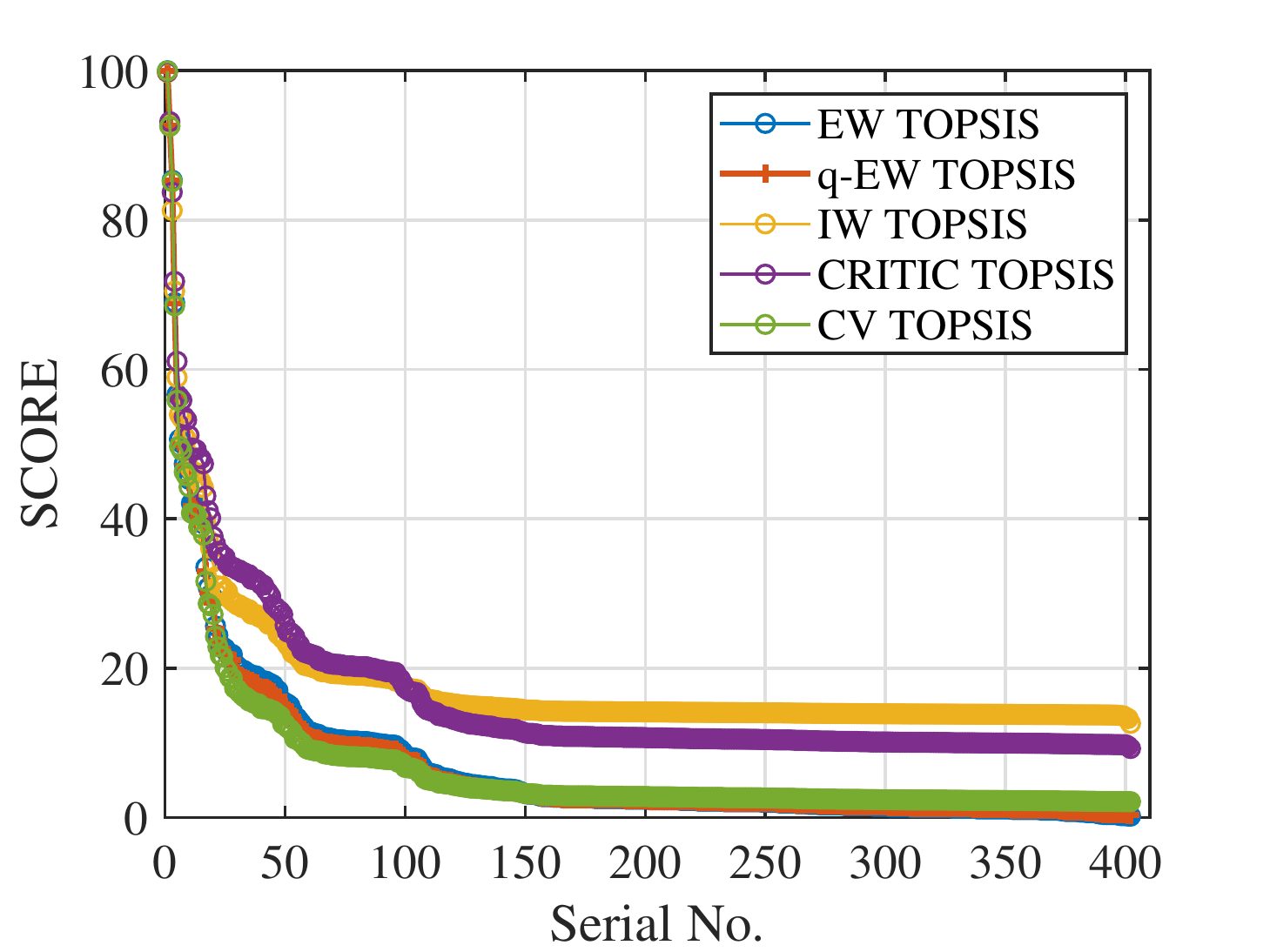}
\caption{TOPSIS score distribution curves for the five objective evaluation methods}
\label{plot-5S}
\end{figure}

It is constructive to plot the distributions of scores obtained in our $q$-EW-TOPSIS model together with the others.
As shown in Figure \ref{plot-5S}, we compare the scores among five different TOPSIS models, namely the classical and non-extensive EW-TOPSIS models, IW-TOPSIS model, CRITIC-TOPSIS model and CV-TOPSIS one.

In details, CV-TOPSIS (Coefficient of Variation TOPSIS), IW-TOPSIS (independent weight coefficient method TOPSIS), and CRITIC-TOPSIS are variants of the TOPSIS method. These variants introduce modifications to the original TOPSIS methodology to address specific issues and improve its performance in decision-making problems.

The CV-TOPSIS (Coefficient of Variation TOPSIS) method introduces the coefficient of variation to consider the stability and relative difference of each indicator data. In the CV-TOPSIS method, the coefficient of variation is calculated for each decision criterion to evaluate the stability of the criterion. The coefficient of variation represents the ratio of the standard deviation of the data set of the criterion to its mean, and the smaller it is, the more stable the criterion. Then, the coefficient of variation is used as a weight and multiplied by each decision criterion's score. By calculating the distance between each decision alternative and the ideal solution, and converting it into a composite score, the alternative is ranked\cite{CV1,CV2,CV3}.

The IW-TOPSIS (Independent Weight Coefficient Method TOPSIS) method is another variation of the TOPSIS method that introduces an independent weight coefficient to consider the importance and relative difference of each criterion data. In the IW-TOPSIS method, the decision maker specifies the independent weight of each decision criterion. The independent weight represents the relative importance of the criterion and determines how the weight is allocated to the decision alternatives. Then, by calculating the distance of each decision alternative from the ideal solution and converting it into a composite score, the alternatives are ranked\cite{IW1,IW2,IW3}.

The CRITIC-TOPSIS (Criteria Weights Determined by Interval Comparison of Rankings) method is an improvement over the TOPSIS method. It introduces the scoring method and the interval comparison method to determine the importance and relative gap of each index data. In the CRITIC-TOPSIS method, the decision maker first specifies the ideal solution for each decision criterion, which is the best score the criterion can reach. The decision maker also specifies the key factors of each decision criterion, which reflect the critical importance of the criterion to the decision outcome. The key factors can be positive or negative and they can respectively enhance or decrease the score of the decision alternative. Then, by calculating the distance between each decision alternative and the ideal solution, converting it into a composite score, the alternatives are ranked\cite{CRITIC1,CRITIC2}.

In general, these three methods are extensions of TOPSIS that improve the accuracy and efficiency of evaluation by expanding the evaluation criteria. They are currently the three most widely used objective evaluation models.

Overall, it is clear from Figure \ref{plot-5S} that our model owns a higher discrimination capability when evaluating suppliers with higher ranking, which is in line with the current industry's demand. 
Additionally, when compared to CV-TOPSIS (green line), it also shows better discrimination in the later stages of the score range. 
Furthermore, the score distribution of our model is not significantly different from the standard entropy weight model, indicating the accuracy of it.

In order to verify the performance of our model when data loss is negligible, we need to examine whether the evaluation method of the gray correlation model is superior to that of the $q$-EW-TOPSIS model. 
Additionally, we want to test whether the model will revert to the weights that can be described by the classical model when data is abundant. 
Therefore, we have chosen a dataset on employee performance evaluation (for instance, the data in this work comes from the open source data of Tempe \cite{dataset2}), with 2,243 participants and 72 evaluation criteria. 
Our goal is to observe the model's performance when data is abundant and to verify its universality in both evaluation and decision-making models.

\begin{figure}[htb]
\centering
\includegraphics[width =0.7\linewidth]{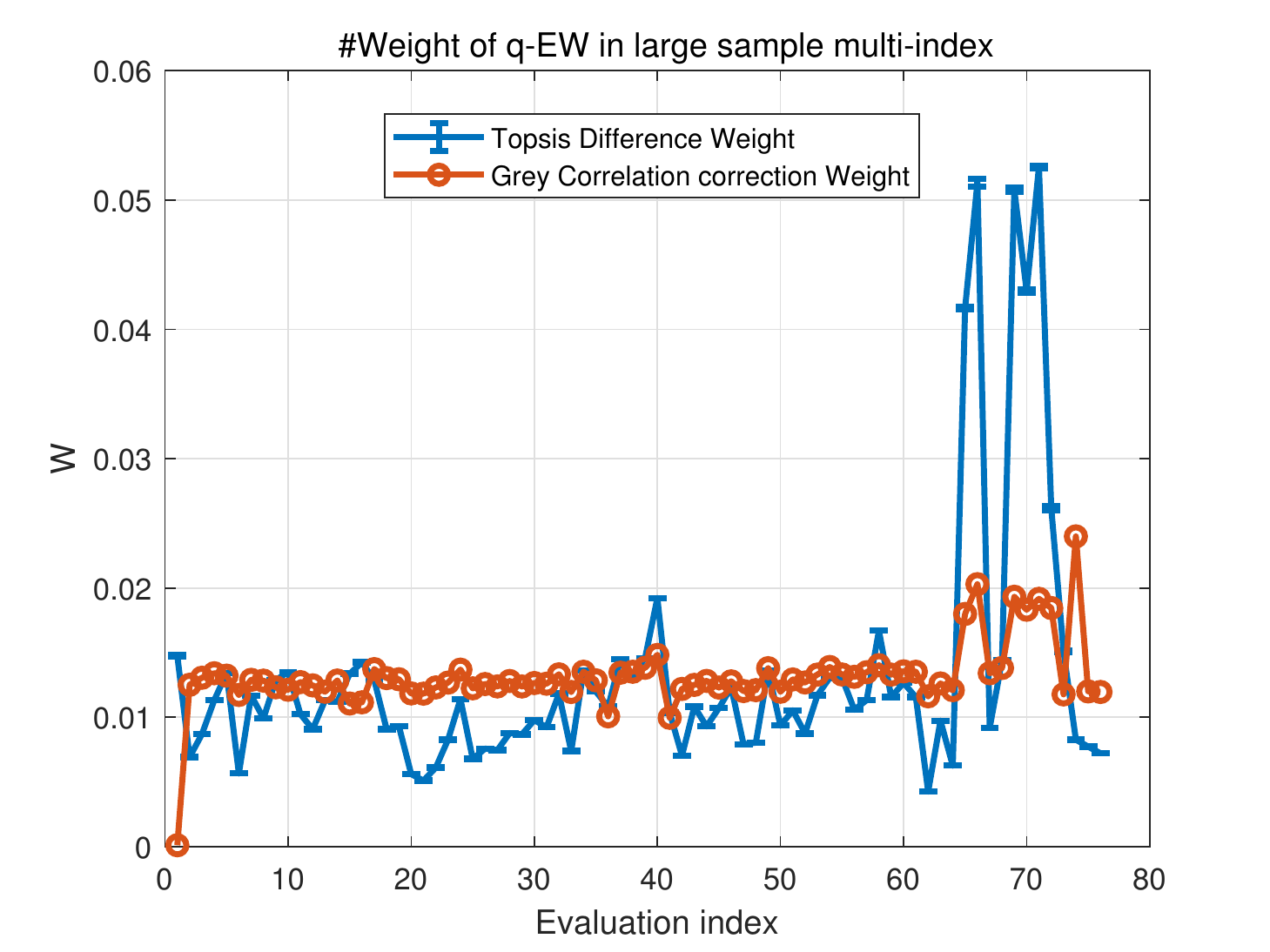}
\caption{Error-bar plot of QEWM versus EWM when the sample size and the number of evaluation indicators are large.}
\label{plot-err}
\end{figure}

As depicted in Figure \ref{plot-err}, our proposed model, $q$-EW-TOPSIS, demonstrates better differentiation among the evaluation indices when applied to a large number of indices. 
This indicates that the model maintains a high level of performance even when the data volume and number of evaluation indices are substantial. 
The mean value of absolute error is found to be $5.02\%$, with a range of $0.0795\% \sim 7.73\%$, all of which fall within an acceptable range. 
Notably, the evaluation results exhibit a higher degree of fluctuation in the middle portion of the dataset but remain relatively stable at the beginning and end. 
This further confirms that $q$-EW-TOPSIS model can effectively address the oversimplification issues encountered in classical models when assessing and ranking the characteristics of suppliers. 
Additionally, this modified method maintains a consistent order among all suppliers.

Furthermore, when the number of evaluation indices, as well as the size of subjects, gets large enough, our model will 
approach to the normal one, which means $q \to 1$.

\section{Summary and Discussion}\label{sec:sec3}

Summarizing, we have generalized the supply capacity evaluation model with respect to the non-extensive statistical entropy.
The traditional EW-TOPSIS model is selected as our baseline and GRA method is used as the correlation feedback to provide the amplitude and direction of model correction, and finally brings out the non-extensive parameter $q$, which will determine the $q$-EW-TOPSIS model in turn.

As discussed above, scores weighted by the classical EW-TOPSIS model distribute too idealized and concentrated, which somehow fails describing the characteristics of suppliers or loses some information accordingly.
For the GRA method, it succeeds in small sample demands or the system that covers few regularities of data request.
Thus it can help deduce the value of $q$, the non-extensive parameter.
With it the $q$-EW-TOPSIS model could be perfectly established.

Compared to the results of the traditional EW-TOPSIS method, the $q$-EW-TOPSIS model maintains its advantages by preserving the sorting order of suppliers, while also expanding the scoring range for increased universality. 
This new model is characterized by its ability to accurately evaluate the weights of indicators, even with smaller sample sizes and weaker rules. 
Unlike traditional models that require larger sample sizes for optimal performance, the $q$-EW-TOPSIS model exhibits superior performance and minimal deviation as a whole.

In this work, it is argued that the use of different indicator weights can capture the ever-changing information in an 
incomplete system. The entropy evaluation method is employed to perform numerical iterations in the system, leading to a 
more stable and closer-to-complete structure. The mutual restriction between indicators is used to overcome the limitations 
of the traditional EW-TOPSIS model in the evaluation of supply capacity, resulting in increased compatibility with the system 
and greater credibility of the results.
By using the entropy evaluation method, it can perform numerical iterations in the system in order to achieve a more stable structure, which is quite closer to the complete one.
In this process, the mutual restriction between different indicators is used to make up for the deficiency of the traditional EW-TOPSIS model itself in the evaluation of supply capacity, making it more compatible with the system and increasing the credibility of the results.

\vspace{0.3cm}

\noindent {\bf Acknowledgments}

\vspace{3mm}

This work has been partly supported by the funding for the Doctoral Research of ECUT Nos. DHBK2019211.
The author Ke-Ming Shen also wants to thank his Tuer for kindly discussions.


\begin{thebibliography}{99}

\bibitem{ref1}
Lima, J. F. R., Osiro, L., $\&$ Carpinetti, L. C. R., ``Supplier selection and order allocation using a multi-objective simulated annealing algorithm," Applied soft computing, vol. 21, pp. 194-209, 2014.

\bibitem{ref2}
George A. Zsidisin, et al., ``Supplier selection: review and analysis of literature from 1994-2004," International Journal of Physical Distribution and Logistics Management, 2004.

\bibitem{ref3}
Vijay R. Kannan, $\&$ Keah Choon Tan, ``Supplier selection: An analytical hierarchy process (AHP) approach," Journal of supply chain management, vol. 38, no. 3, pp. 11-21, 2002.

\bibitem{ref4}
Narges Banaeian, et al., ``Supplier selection using a hybrid MCDM method based on grey relational analysis and DEMATEL," Computers $\&$ Operations Research, vol. 89, pp. 337-347, 2018.

\bibitem{AHP}
Thomas L. Saaty, ``The Analytic Hierarchy Process," Springer, Berlin, Heidelberg, pp. 109-121, 1988.

\bibitem{AHP1}
Thomas L. Saaty, ``The Hierarchon: A Dictionary of Hierarchies," International Journal of Services Sciences, vol. 1, no. 1, pp. 83-98, 2008.

\bibitem{cem}
Mithat Zeydan, Cuneyt Colpan, and Cemal Cobanoglu, ``A novel MCDM approach for supplier selection: Interval-valued 2-tuple linguistic grey relational analysis," Expert systems with applications, vol. 38, no. 3, pp. 2741-2751, 2011.

\bibitem{data}
Abdullah Aldamak and Saeed Zolfaghari, ``A new hybrid MCDM model for supplier selection using the combination of fuzzy TOPSIS and entropy," Measurement, vol. 106, pp. 161-172, 2017.

\bibitem{BP} 
Celebi, Dilay, and Demet Bayraktar. ``An integrated neural network and data envelopment analysis for supplier evaluation under incomplete information." Expert Systems with Applications 35.4 : 1698-1710, 2008.

\bibitem{BP1} 
Quan, Quan, and Zhongqiang Zhang. ``Supply capability evaluation of intelligent manufacturing enterprises based on improved BP neural network." Journal of Mathematics 2022: 1-8, 2022.

\bibitem{BP2} 
Liu, Li, and Wenxue Ran. ``Research on supply chain partner selection method based on BP neural network." Neural Computing and Applications 32: 1543-1553, 2020.

\bibitem{TOPSIS0}
Velasquez, M. $\&$ Hester, P.T., ``International journal of operations research." International Journal of Operations Research, 10(2), 56-66, 2013.

\bibitem{TOPSIS}
Behzadian, Majid, et al. ``A state-of the-art survey of TOPSIS applications." Expert Systems with applications 39.17: 13051-13069, 2012.

\bibitem{EWM}
Xing, Yahong, et al. ``Evaluation System of Distribution Network Admission to Roof Distributed Photovoltaic Based on AHP-EW-TOPSIS." 2021 IEEE 5th Conference on Energy Internet and Energy System Integration (EI2). IEEE, 2021.

\bibitem{Renyi}
A. Renyi. Probability Theory. North Holland, Amsterdam, 1970.

\bibitem{TA}
Tsallis, C., ``Possible generalization of Boltzmann-Gibbs statistics," J. Stat. Phys., vol. 52, pp. 479, 1988.

\bibitem{Abe}
Abe, Sumiyoshi. ``A note on the q-deformation-theoretic aspect of the generalized entropies in nonextensive physics." Physics Letters A 224.6: 326-330, 1997.

\bibitem{Kaniadakis}
G. Kaniadakis, ``Non-linear kinetics underlying generalized statistics," Physica A, vol. 296, pp. 405, 2001.
 
\bibitem{ref6}
Hwang Ching-Lai and Kwangsun Yoon, Multiple attribute decision making. Springer, Berlin, Heidelberg.

\bibitem{ref8}
Zhou, L., $\&$ Li, X. (n.d.). ``Supplier selection based on grey system theory and improved entropy weight method." The 8th China Intelligent Computing Conference and the 3rd Annual Conference of the China Branch of the International Electronic Commerce Federation.

\bibitem{GRA}
Deng, J.L., ``The Introduction to Grey System Theory,'' The Journal of Grey System, Vol. 1, No. 1, pp.1-24, 1989

\bibitem{GRA1}
Luthra, Sunil, et al. ``An integrated framework for sustainable supplier selection and evaluation in supply chains." Journal of cleaner production 140: 1686-1698, 2017.

\bibitem{GRA2}
Rajesh, R., and V. Ravi. ``Supplier selection in resilient supply chains: a grey relational analysis approach." Journal of Cleaner Production 86: 343-359, 2015.

\bibitem{dataset2}
https://data.tempe.gov/datasets/city-of-tempe-2016-employee-survey-data/explore

\bibitem{CV1}
T. Sun, J. Wang and C. Wang, ``TOPSIS threat assessment based on the coefficient of variation," 2020 39th Chinese Control Conference (CCC), Shenyang, China, pp. 175-179, 2020.

\bibitem{CV2}
Vavrek, Roman, and Jana Chovancová. ``Assessment of economic and environmental energy performance of EU countries using CV-TOPSIS technique." Ecological Indicators 106: 105519, 2019.

\bibitem{CV3}
H. Su and D. Z. Du, ``The Improved Coefficient of Variation TOPSIS Model and its Application," Measurement, vol. 72, pp. 174-183, 2015.

\bibitem{IW1}
Hai-dong, W. U., et al. ``Evaluation of Driving Behavior Economy Based on Big Data of New Energy Bus." China Journal of Highway and Transport 35.3 : 177, 2022.

\bibitem{IW2}
Scientific Platform Serving for Statistics Professional 2021. SPSSPRO

\bibitem{IW3}
Jiang, Tingting, et al. ``Strategy of Energy Conservation and Emission Reduction in Residential Building Sector: A Case Study of Jiangsu Province, China." Journal of Environmental and Public Health 2023, 2023.

\bibitem{CRITIC1}
Mohamadghasemi, A., Hadi$\-$Vencheh, A., $\&$ Hosseinzadeh Lotfi, F., The multiobjective stochastic CRITIC$\-$TOPSIS approach for solving the shipboard crane selection problem. International Journal of Intelligent Systems, 35(10), 1570-1598,2020.

\bibitem{CRITIC2}
Rostamzadeh, Reza, et al. ``Evaluation of sustainable supply chain risk management using an integrated fuzzy TOPSIS-CRITIC approach." Journal of Cleaner Production 175: 651-669, 2018.

\bibitem{dataset}
Chinese Society for Industrial and Applied Mathematics.,The ordering and transportation of raw materials for a manufacturing enterprise.question C, 2021.
$http://www.mcm.edu.cn/html\_cn/node/4d73a36cc88b35bd4883c276afe39d89.html$

\end{thebibliography}
\end{document}